\documentclass[12pt]{article}

\usepackage{hyperref}

\usepackage{scicite}

\usepackage{times}

\usepackage{graphicx}

\usepackage{xurl}

\usepackage{booktabs}

\newenvironment{sciabstract}{%
\begin{quote} \bf}
{\end{quote}}

\title{Validating daily social media macroscopes of emotions}

\author
{Max Pellert,$^{1,2,3,4\ast}$ Hannah Metzler,$^{2,3,4,5}$ Michael Matzenberger,$^{6}$ David Garcia$^{2,3,4}$\\
\\
\normalsize{$^{1}$Sony Computer Science Laboratories,}\\
\normalsize{6, Rue Amyot, 75005 Paris, France}\\
\normalsize{$^{2}$Institute of Interactive Systems and Data Science,}\\
\normalsize{Graz University of Technology,}\\
\normalsize{Inffeldgasse 16C, 8010 Graz, Austria}\\
\normalsize{$^{3}$Complexity Science Hub Vienna,}\\
\normalsize{Josefstädter Straße 39, 1080 Vienna, Austria}\\
\normalsize{$^{4}$Section for Science of Complex Systems, Center for Medical Statistics,}\\
\normalsize{Informatics and Intelligent Systems, Medical University of Vienna,}\\
\normalsize{Spitalgasse 23, 1090 Vienna, Austria}\\
\normalsize{$^{5}$Institute of Globally Distributed Open Research and Education}\\
\normalsize{$^{6}$Der Standard,}\\
\normalsize{Vordere Zollamtsstraße 13, 1030 Vienna, Austria}\\
\\
\normalsize{$^\ast$To whom correspondence should be addressed; E-mail: pellert@csh.ac.at.}
}

\date{}

\begin{document} 


\baselineskip24pt


\maketitle


\begin{sciabstract}
Measuring sentiment in social media text has become an important practice in studying emotions at the macroscopic level. However, this approach can suffer from methodological issues like sampling biases and measurement errors. To date, it has not been validated if social media sentiment can actually measure the temporal dynamics of mood and emotions aggregated at the level of communities. We ran a large-scale survey at an online newspaper to gather daily mood self-reports from its users, and compare these with aggregated results of sentiment analysis of user discussions. We find strong correlations between text analysis results and levels of self-reported feelings, as well as between inter-day changes of both measurements. We replicate these results using sentiment data from Twitter. We show that a combination of supervised text analysis methods based on novel deep learning architectures and unsupervised dictionary-based methods have high agreement with the time series of aggregated mood measured with self-reports. Our findings indicate that macro level dynamics of feelings expressed on an online platform can be tracked with social media text, especially in situations of high mood variability.
\end{sciabstract}

\section*{Introduction}

User generated text from social media has become an important data source to analyze expressed mood and emotions at large scales and high temporal resolutions, for example to study seasonal mood oscillations \cite{golder2011diurnal}, emotional responses to traumatic events \cite{garcia2019collective}, the effect of pollution on happiness \cite{zheng2019air}, and the role of climate change in suicide and depression \cite{burke2018higher}.
Despite these promising applications, using social media text to measure emotion aggregates can suffer a series of methodological issues typical of studies of this kind of \emph{found data} \cite{ruths2014social,olteanu2019social,sen2019total}. 
Common validity threats are measurement error in sentiment analysis tools and the performative behavior of social media users due to platform effects or community norms. Sampling biases can generate a mismatch between users that produce text and a target group that might include silent individuals.

The validation of sentiment analysis methods has focused on micro level measurement accuracy at the individual post level \cite{ribeiro2016sentibench}.
Recent work has assessed the measurement validity also at the individual person level, using historical records of text from a user. This has revealed low to moderate correlations between aggregates of sentiment produced by an individual over a period of time and emotion questionnaires \cite{beasley2015emotional,krossDoesCountingEmotion2019}. At the group level, static measurements of social media sentiment are only moderately correlated with affective well-being and life-satisfaction across regions  \cite{jaidkaEstimatingGeographicSubjective2020}. These earlier findings highlight the limits of static aggregations of sentiment to measure concepts like life satisfaction that are only slowly changing over time. However, it is still an open question if analyses of social media text can shed light on \textit{faster} phenomena, for example core emotional experiences, when we stick to aggregating individual signals to a community of interest and observe variation over time.

Here, we address this research gap by testing whether social media text sentiment tracks the macro level dynamics of emotions with daily resolution in an online community. We study the convergence validity of two approaches to study \textit{emotions} at scale: \textit{sentiment} aggregates from social media text and \textit{mood} self-report frequencies in a survey. For 20 days, we collected 268,128 emotion self-reports through a survey in an Austrian online newspaper. During the same period, we retrieved text data from user discussions on the same platform, including 452,013 posts in our analysis using our pre-existing Austrian social media monitor \cite{pellertDashboardSentimentAustrian2020a}. To replicate our results with a second dataset, we conducted a pre-registered analysis of 635,185 tweets by Austrian Twitter users. We applied two off-the-shelf German sentiment analysis tools on the text data: a state-of-the-art supervised tool based on deep learning (German Sentiment, GS, \cite{guhr-EtAl:2020:LREC}) and a popular dictionary method based on expert word lists (Linguistic Inquiry and Word Count, LIWC, \cite{wolf_computergestutzte_2008}). Our results strongly support the assumption that social media sentiment can reflect both mean levels and changes of self-reported emotions in explicit daily surveys. We additionally analyze positive and negative components of the sentiment analysis methods to provide further methodological insight on how to measure time series of experienced mood in online communities.

\section*{Results}

We measured the time series of experienced mood as the fraction of self-reported positive feelings over the total of self-reports in a day. We coupled this with a daily sentiment aggregate based on text in the platform's forum (derstandard.at), namely the average of the GS and LIWC scores (see Sentiment analysis and Text data in Materials and Methods for more details).

\begin{figure*}[t]
\centering
\includegraphics[width=0.98\linewidth]{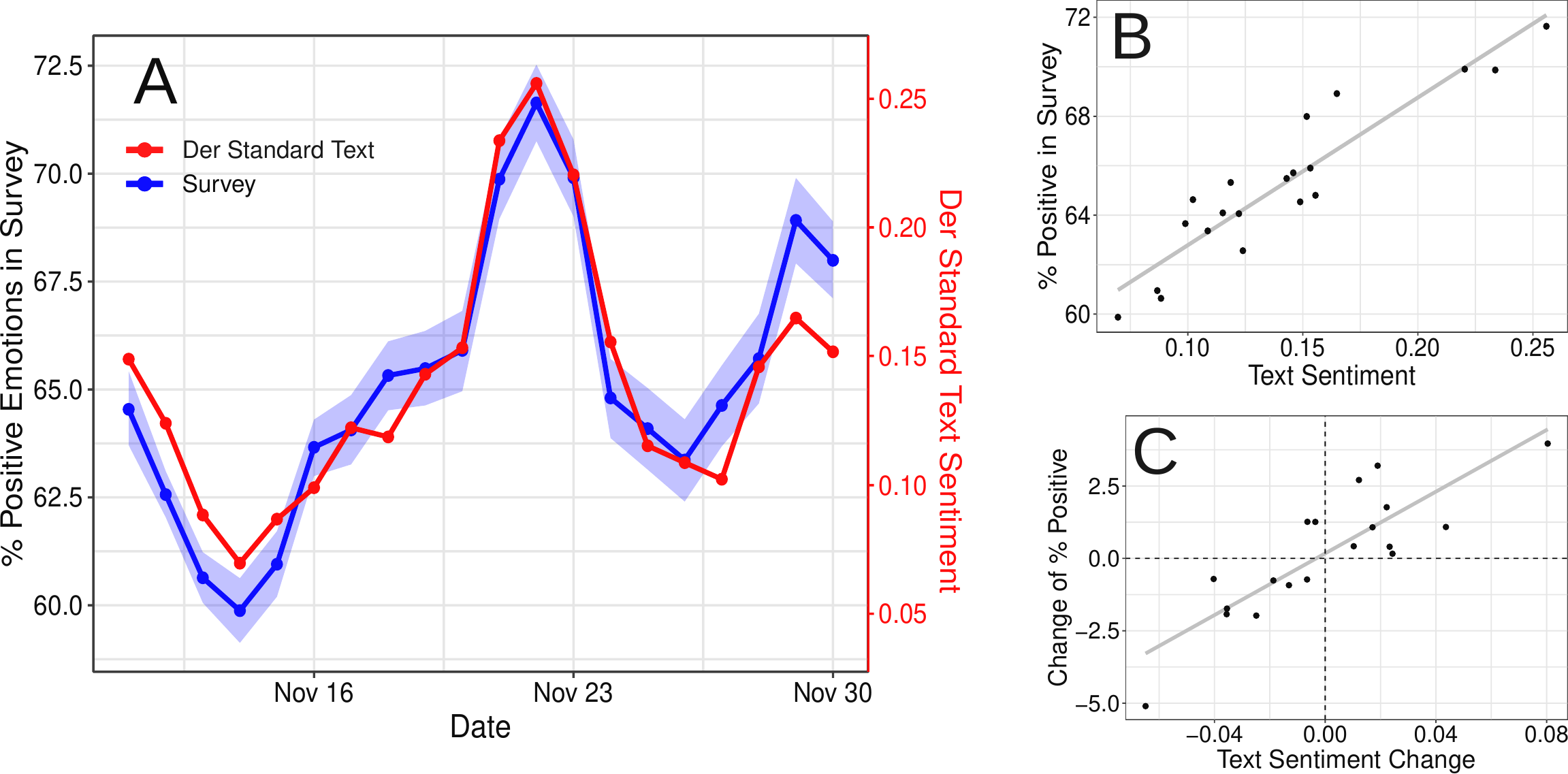}
\caption{Panel A: Time series of the daily percentage of positive mood reported in the survey and the aggregated sentiment of user-generated text on derstandard.at. The shaded blue area corresponds to 95\% bootstrapped confidence intervals. Panel B: Scatterplot of text sentiment and survey responses with regression line. Panel C: Scatterplot of the daily \textit{changes} in both text sentiment and survey responses compared to the previous day, with regression line. }\label{fig:twolineplot_derstandard}
\end{figure*}

Figure \ref{fig:twolineplot_derstandard} shows the time series of the fraction of responses that report a positive mood and the text sentiment aggregate from the Der Standard forum. The Pearson correlation coefficient between both measurements is $0.93$ ($[0.82,0.97],p<10^{-8}$), indicating a very strong positive correlation with daily resolution over a period of almost three weeks. The regression line in the scatter plot on Panel B in Figure \ref{fig:twolineplot_derstandard} confirms the relationship. A linear model with Heteroskedasticity and Autocorrelation Consistent (HAC) estimates shows the same robust effect. The model has an adjusted R-squared of $0.852$, with a coefficient $\hat{\beta}=0.597$ ($[0.465,0.728],p<10^{-7}$) for the unscaled average of sentiment aggregates. The text sentiment aggregate can explain 85\% of the variance in the daily proportion of positive mood.

We additionally tested if \textit{changes} in the text sentiment aggregate can approximate daily changes in the proportion of positive mood in the survey compared to the previous day. A similar regression model as before yields a coefficient of $\hat{\beta}=0.533$ ($[0.390,0.675],p<10^{-16}$) for changes in the text sentiment aggregate and an adjusted R-squared of $0.704$ (Panel C of Figure \ref{fig:twolineplot_derstandard}). This model has a non-significant intercept of $0.002$ ($[-0.002,0.005],p=0.29$) showing that, in addition to explaining 70\% of the variance in emotion changes at the macro level, the model's prediction of trend in mood changes is not significantly biased.

To test the robustness of our results as well as their generalizability to a different platform, we pre-registered a replication of our analysis using 515,187 tweets by Austrian Twitter users in the survey period instead of Der Standard forum posts (\url{https://aspredicted.org/blind.php?x=vb3gp2)} (see Methods for more details on sample size and selection criteria). The correlation coefficient between the survey and sentiment on Twitter is positive and significant ($0.63$ $[0.26,0.84],p<0.003$), confirming our pre-registered hypothesis and the robustness and generalizability of the results. Although it is somewhat lower than the correlation of the survey with text sentiment from the Der Standard forum, a coefficient above 0.6 is still sizeable, especially given that the survey and the postings now come from different platforms. Following our pre-registration, we filtered out accounts tagged as "organisational" by the aggregation service Brandwatch (formerly known as Crimson Hexagon) and accounts with less than 100 followers or more than 5000 followers. If we relax this criterion to include accounts with up to 100000 followers as in our previous study \cite{metzler_collective_2021}, the correlation increases to $0.71$ ($[0.39, 0.88],p<0.0005$). This suggests that influential accounts are also relevant to calculate sentiment aggregates, as central individuals in the Twitter social network might be serving as early sensors of sentiment shifts \cite{galesicHumanSocialSensing2021,garcia-herranzUsingFriendsSensors2014,garcia2021social}.

\begin{figure*}[t]
\centering
\includegraphics[width=0.98\linewidth]{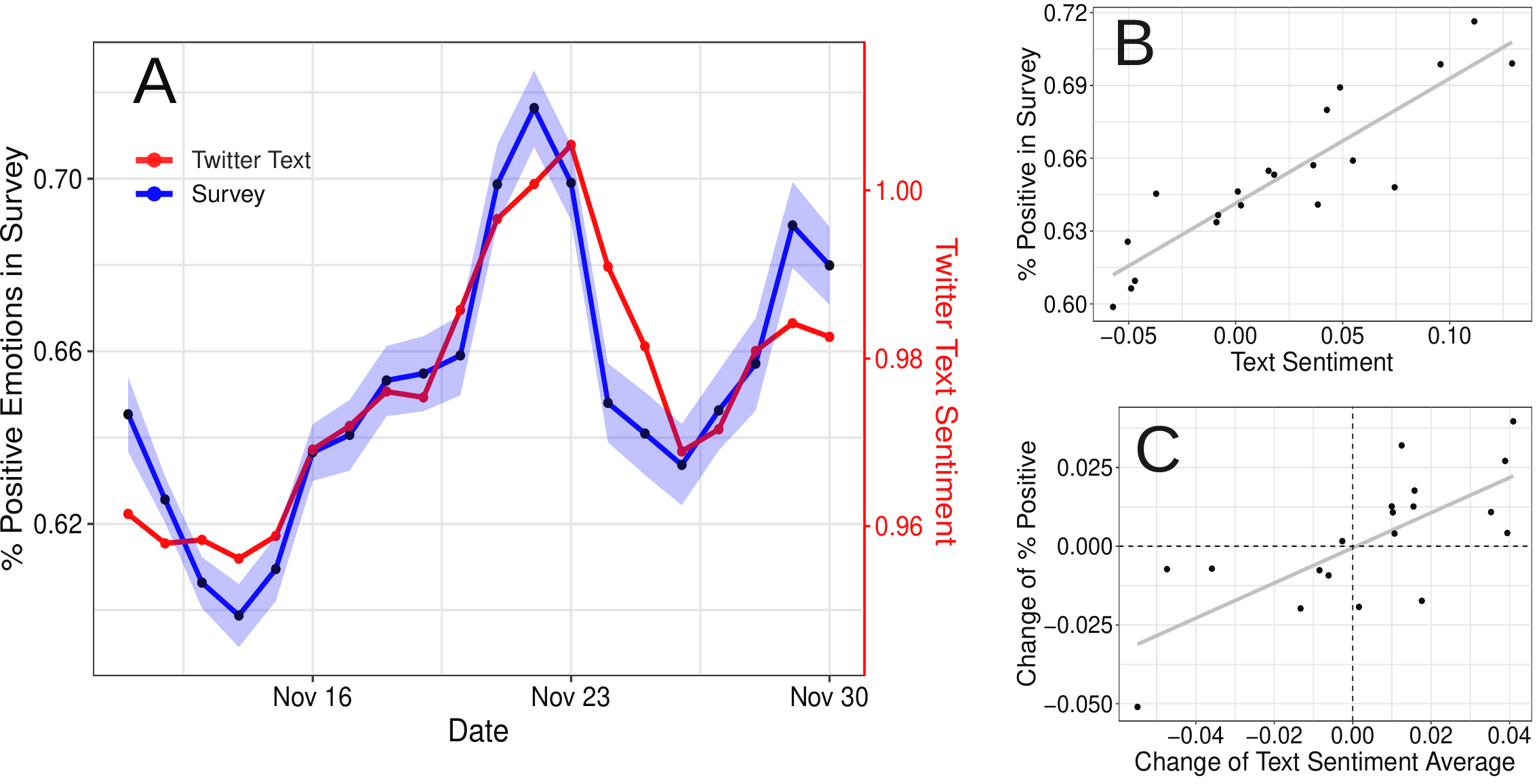}
\caption{Panel A: Time series of the daily percentage of positive feelings reported in the survey and the aggregated sentiment of user-generated text on Twitter in Austria. The shaded blue area corresponds to 95\% bootstrapped confidence intervals. Panel B: Scatterplot of text sentiment and survey responses with regression line. Panel C: Scatterplot of the daily \textit{changes} in both text sentiment and survey responses compared to the previous day, with regression line. }\label{fig:twolineplot_twitter}
\end{figure*}

Beyond our pre-registered hypothesis, we found that the Twitter sentiment signal is lagged by a day compared to the mood survey. Figure S2 shows the data with a shift of one day in comparison to no shift. Correcting this by shifting by one day yields a correlation coefficient of $0.90 ([0.75,0.96],p<10^{-6})$. We see one explanation for this: The newspaper articles and discussions of their contents likely capture immediate reactions to events, while reaching the wider audience of Twitter takes longer. Panel A of Figure \ref{fig:twolineplot_twitter} shows that the survey and Twitter time series closely track each other, and the regression line in Panel B confirms this (coefficient $\hat{\beta}=0.516$ ($[0.379,0.654],p<10^{-16}$ with an adjusted R-squared of $0.791$). Again, changes of both variables (Panel C) also have a strong relationship, indicated by a regression coefficient of $\hat{\beta}=0.557$ ($[0.296,0.819],p<10^{-16}$), an adjusted R-squared of $0.501$ and a non-significant intercept of $-0.00051$ ($[-0.009,0.008],p=0.90$). For the remaining analysis, we build on this model with a lag of one day to understand the best case of how Twitter sentiment can explain the survey. 

We further explored which components of sentiment analysis are the most informative when estimating the daily proportion of self-reported positive feelings. Table \ref{tab:cor} shows the correlation of positive feelings with the aggregated values (positive minus negative emotions, averaged across the LIWC and GS measure) as well as the positive and negative components of both sentiment analysis methods separately. All variables were rescaled through a Z-transformation for both Der Standard and Twitter postings. The positive component of GS has a high correlation with the proportion of self-reported positive mood.
The positive component of LIWC also has a positive, but somewhat lower coefficient for both Der Standard and Twitter.
Additionally, the LIWC and the GS positive components are strongly correlated with each other on both platforms (Der Standard: $\rho=0.94$ $[0.85,0.98],p<10^{-8}$, Twitter: $\rho=0.90$ $[0.75,0.96],p<10^{-6}$), indicating convergent validity of the two methodologically distinct measures of positive emotions from text when aggregating as daily frequencies of emotional expressions. The analysis of LIWC scores shows inconsistencies: The negative component of LIWC does not correlate with the proportion of positive mood in the survey for Der Standard. Yet, LIWC negative is informative for Twitter, with a significant negative correlation coefficient. Overall, comparing machine-learning and dictionary-based methods shows that the supervised classifier shows more consistent performance and generally higher point estimates. Yet,   confidence intervals overlap, and negative LIWC beats GS on Twitter data. Combining both methods for Der Standard adds a small increase to the already strong correlations of the supervised classifier alone. Taken together, it thus seems that both methods contribute unique variation for explaining self-reported feelings. 

\begin{table}[hb!]
\centering
\caption{Correlation of positive mood in the survey with text sentiment measures on both platforms (Der Standard and Twitter). The table presents sentiment aggregates (positive minus negative emotions), as well as positive and negative components separately. LIWC+GS indicates the average across both sentiment analysis methods, all other lines present aggregates or components separately for each method. Shift 1 denotes a shift of one day, where survey values precede Twitter values.}
\bigskip
\label{tab:cor}
\begin{tabular}{rlll}
  \toprule
 & Der Standard (No shift) & Twitter (Shift 1) & Twitter (No shift) \\ 
  \midrule
LIWC+GS & 0.93 [0.82,0.97] & 0.90 [0.75,0.96] & 0.71 [0.39,0.88] \\ 
  LIWC & 0.74 [0.44,0.89] & 0.85 [0.65,0.94] & 0.66 [0.31,0.85] \\ 
  LIWC pos & 0.81 [0.56,0.92] & 0.80 [0.56,0.92] & 0.60 [0.22,0.83] \\ 
  LIWC neg & 0.03 [-0.42,0.46] & -0.74 [-0.89,-0.43] & -0.63 [-0.84,-0.26] \\ 
  GS & 0.91 [0.78,0.96] & 0.91 [0.79,0.96] & 0.73 [0.43,0.89] \\ 
  GS pos & 0.89 [0.75,0.96] & 0.91 [0.79,0.97] & 0.80 [0.54,0.92] \\ 
  GS neg & -0.57 [-0.81,-0.18] & -0.39 [-0.71,0.06] & -0.17 [-0.57,0.3] \\ 
   \bottomrule
\end{tabular}
\end{table}

The strong relationship that we found between the signals of the survey, Der Standard postings and tweets opens up the possibility to measure emotion aggregates through social media text when survey data is not available. This raises the question how these three different affective measures correlate with external events that drive the emotions of a community. In the following, we test if our social media text measures provide comparable correlations with the number of new COVID-19 cases to self-reports in surveys. The survey period in November 2020 falls within the build-up of the plateau of COVID-19 cases in the second wave of the pandemic in autumn 2020 in Austria, providing an ideal time frame to test this hypothesis. The importance of the topic in public discussion makes new COVID-19 cases a relevant external variable that previous research has linked to emotional experiences in the population as a whole \cite{metzler_collective_2021}. We retrieved COVID-19 case data from Our World in Data \cite{roser2020coronavirus} for the period that overlaps with the survey. The survey and both aggregate Twitter sentiment measures (with and without shift) correlate significantly with the number of new cases for the corresponding day (Table \ref{tab:casecor}). Der Standard sentiment shows a weaker correlation and is not significant. Figure S3 confirms the relationships with scatter plots. Furthermore, we tested if this relationship with new COVID cases significantly differs when correlating survey feelings compared to text sentiment measures (Table S1 and Figure S4 in SI): The correlation obtained with both sentiment analysis methods for Twitter data does not significantly differ from the correlation with survey data.

\begin{table}[hb!]
\centering
\caption{Correlation of survey, aggregate Twitter sentiment and aggregate Der Standard sentiment with the number of new COVID-19 cases. Figure S3 shows scatter plots for each of the variables and new COVID-19 cases.}
\bigskip
\label{tab:casecor}
\begin{tabular}{rl}
  \toprule
 & New Cases \\ 
  \midrule
Twitter (Aggregate Shift 1) & -0.60 [-0.82,-0.21] \\ 
  Twitter (Aggregate No Shift) & -0.57 [-0.81,-0.17] \\ 
  Survey & -0.53 [-0.79,-0.12] \\ 
  Der Standard (Aggregate) & -0.33 [-0.68,0.13] \\ 
   \bottomrule
\end{tabular}
\end{table}

We built on the strong correlation of sentiment of both platforms with the survey to study their relationship also outside of the survey time frame: Figure \ref{fig:extsignalanalysis} shows the time series of the two platform's sentiment signal before, during and after the survey. For the period between 2020-09-15 until 2021-12-30, we find a significant positive correlation of $\rho=0.53$ $[0.38,0.65],p<10^{-16}$. This additional analysis shows that the relationship of expressed emotions on the two platforms is stable also over longer time periods.

\section*{Discussion}

This study compared online newspaper readers' self-reported feelings with sentiment analysed in postings on two social media platforms. The results show that the sentiment contained in text of postings in the online discussion forum on the newspaper's website tracks the daily frequency of self-reported feelings in the survey. These results generalize across social media platforms, as a pre-registered replication shows similar correlations when using Twitter text instead of postings in the newspaper forum. Despite the methodological challenges in studying affective states with social media text, this provides evidence that the aggregate of sentiment analysis of social media text can be used to measure macro level mood. We find strong relationships with both levels of emotions and inter-day changes, showing that social media sentiment indeed tracks macro-level mood dynamics with daily resolution. This differs from previous studies reporting only low positive (or even negative) correlations between self-reported well-being indicators and long-term dictionary-based positive sentiment aggregates across US regions \cite{jaidkaEstimatingGeographicSubjective2020}. In contrast, we find that dictionary as well as machine-learning based methods track short-term affective states, as opposed to more long-term concepts such as life satisfaction. Additionally, we find that adding indicators based on unsupervised dictionary methods increases the already high agreement of supervised machine-learning methods with the survey measurement of macro level emotions. This suggests that dictionaries generated by experts, while being less exhaustive than the large models of supervised methods, may include terms that are not discovered or not attributed adequate importance in the training phase of supervised models.

Other methodological issues of social media research remain unsolved, for example if social media sentiment measures the emotions of the wider population beyond an online community. However, all readers of the online newspaper Der Standard were prompted to anonymously fill in the survey. To not only consider active users of the forum with a registered account lowers the barrier for participation substantially. Our results provide first evidence that postings by active users can also reflect emotions of silent individuals in an online community to a very high degree. A potential influence on this relationship points in the direction of emotional contagion \cite{goldenbergDigitalEmotionContagion2020c}. Emotions can spread between registered forum user through their postings to unregistered users. \textit{Digital} emotional contagion was indeed observed on social media both in studies of randoms sample of users \cite{ferraraMeasuringEmotionalContagion2015} as well as of users that displayed synchronized emotional sharing in the aftermath of a terrorist attack \cite{garciaCollectiveEmotionsSocial2019a}.

\begin{figure*}[!hb]
\centering
\includegraphics[width=0.8\linewidth]{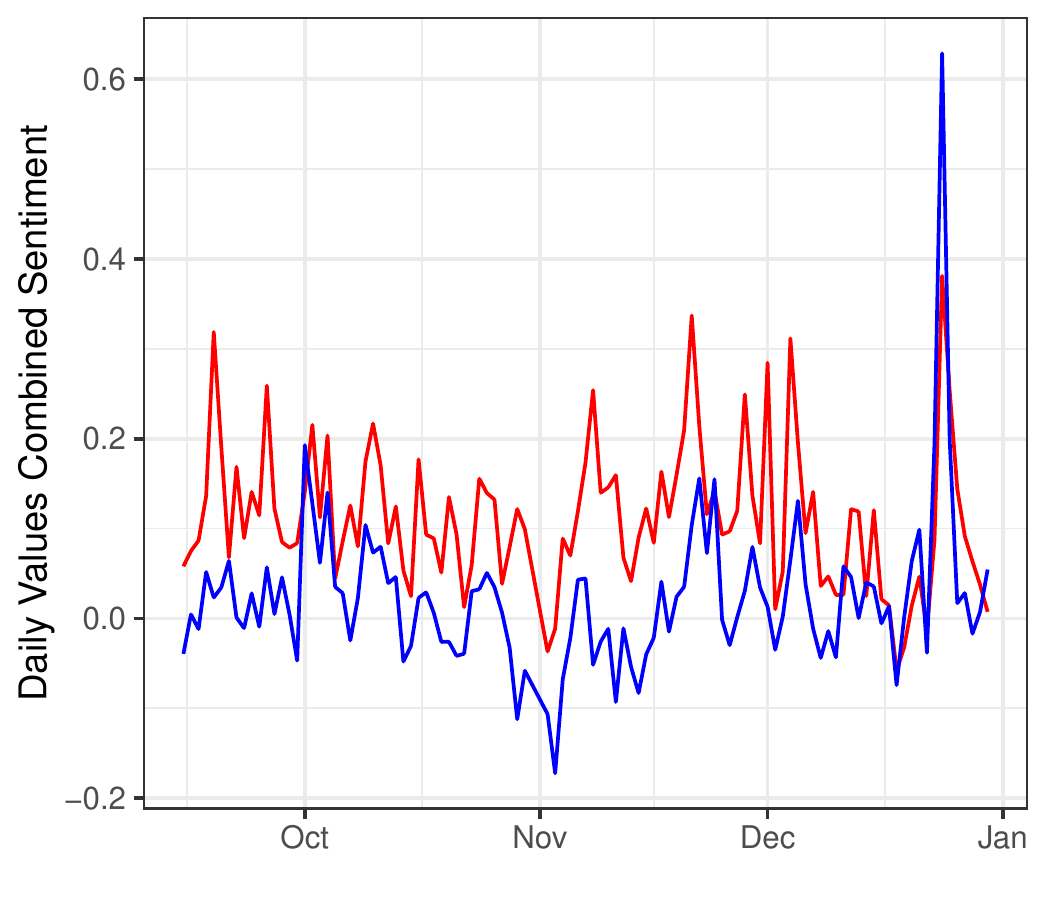}
\caption{Time series of the aggregate (LIWC + GS) sentiment measure for Twitter (blue) and Der Standard (red) covering the time period between 2020-09-15 and 2021-12-30.}\label{fig:extanalysis}.
\label{fig:extsignalanalysis}
\end{figure*}

Beyond this, we show that our findings generalise to the online ecosystem of Twitter users, who most likely did not participate in the survey on Der Standard. With Der Standard as well as Twitter, we study two different online ecosystems whose users are based in the same geographical area and have similar demographic characteristics. There could potentially be a certain overlap in their user bases, but precise estimates are currently not possible (for privacy reasons, Der Standard users and survey respondents cannot be individually identified). Another limitation concerns the relatively short time period during which the survey was online, which was determined by administrative decisions at Der Standard. Future research should look for opportunities to distribute or access more long-term surveys, to test if these results extend to longer time periods. To address the question of longer term stability with the data that we have available, we compared the two text sentiment signals over an extended time period. We showed that their strong relationship is also present for several months before, during and after the survey period (Figure \ref{fig:extsignalanalysis}).

When comparing the components of sentiment analysis methods, we find that positive sentiment is generally more informative than negative sentiment when analysing daily data from these German-speaking social media platforms.
The negative affect dictionary measure does not significantly correlate in the Der Standard dataset while performing better, but still worse than positive affect, on Twitter. In contrast, in previous Twitter studies in English, LIWC negative affect signals were more informative than positive affect signals when studying well-being across regions \cite{jaidkaEstimatingGeographicSubjective2020}. A concurrent study tracks weekly emotions in the UK and also finds stronger correlations with negative emotions measured with English LIWC \cite{garcia2021social}. Our case of the German language analysis on Der Standard differs from those results, pointing to a potentially missing methodological link that connects which kind of sentiment captures which emotions or sentiments on social media text. A word shift graph \cite{gallagherGeneralizedWordShift2021} (Figure S14) shows the higher prevalence of Austrian German dialect words on Der Standard as the biggest difference between the two platforms' text corpora during the survey period. We could assume that users use dialect words more often to express negative than positive affect (to swear or to express general discontent for example). This may explain the worse performance for dictionary methods such as German negative LIWC, and suggests explicitly including such dialect expressions (with no standard spelling) in the dictionary is warranted. GS, on the other hand, is trained on large amounts of "in-the-wild" texts from the internet and may already have encountered such non-standard expressions, or be able to infer their meaning from the context.
Still, we also noticed generally weaker correlations for negative than positive sentiment with the GS method both on Der Standard and Twitter. A possible explanation for this discrepancy is an asymmetry in measurement error when detecting positive versus negative sentiment, which has been shown for English to be potentially different across various social media \cite{ribeiro2016sentibench} and especially challenging in political discussions \cite{thelwall2014heart}. As emotional expression has a tendency to be positive  \cite{boucherPollyannaHypothesis1969,garcia2012positive}, the large-scale training corpora in German \cite{ortiz-suarez-etal-2020-monolingual} for supervised methods like GS might have substantially more positive than negative text to learn from. In line with the current view in NLP research of "more is better" \cite{brownLanguageModelsAre2020}, a possible avenue to improve supervised sentiment analysis methods is to include additional negative texts or to generate balanced samples with respect to sentiment to improve negative sentiment detection.

Our results do not imply that all sentiment analyses on any social media platform will reflect macro level emotions. However, we show that social media data can reflect macro emotions, in particular for short-term emotional states, and that this can be validated against survey data. Aggregates of social media text analysis can serve as \emph{macroscopes} which combine measurements that may be noisy at the level of individuals or posts, but, when aggregated across thousands of posts per day, can provide a valid signal that strongly correlates with the results of standard social scientific methods like surveys.

\section*{Methods}

Part of this analysis was pre-registered at \url{https://aspredicted.org/blind.php?x=vb3gp2}. Specifically, we pre-registered the methodology we previously developed for analysing sentiment in text from Der Standard, and then tested the robustness of the results by repeating the same analysis with text from Twitter. Figures S2-S6 provide information on demographics of users on Der Standard and Austrian Twitter. Users of the two platforms tend to be more often male, younger, more highly educated and more often from Vienna or Upper Austria than respondents of a representative survey in Austria \cite{niederkrotenthaler_mental_2022}. We average over waves 4-6 of that survey, the waves in which questions about Der Standard and Twitter were included. Of a total of 3002 respondents, 533 ($\sim 18\%$) report having a Twitter account and 200 ($\sim 7\%$) report having an account on Der Standard.

\subsection*{Survey data}

The survey was displayed after the text of all articles in the Austrian online newspaper derstandard.at between November 11th and 30th, 2020 (for an example see Figure S1). Der Standard is one of the major newspapers in Austria and its online community is highly active, with almost 57 million visits in November 2020. The headline of the survey was "How was your last day" ("Wie war der letzte Tag?") and the question displayed to respondents was "If you think back to the previous day, do you have a positive or a negative feeling?" ("Wenn Sie an den gestrigen Tag denken, haben Sie ein positives oder negatives Gefühl?"). Respondents had the following choices: "good" ("Gut"), "somewhat good" ("Eher gut"), "somewhat bad" ("Eher schlecht") and "bad" ("Schlecht"). This retrospective assessment is known as the day reconstruction method that was shown to reduce errors and biases of recall \cite{kahnemanSurveyMethodCharacterizing2004}. In comparison to experience sampling methods that rely on repeatedly probing in real-time, the day reconstruction method is non-disruptive, places less burden on respondents, and provides an assessment of the experience of whole days instead of momentary snapshots. It has been used in research to study for example the experience of pain \cite{kruegerAssessmentPainCommunitybased2008}, the relationship of socio-economic status to the prevalence of a number of common illnesses \cite{stoneSocioeconomicGradientDaily2010}, the influence of age on psychological well-being \cite{stoneSnapshotAgeDistribution2010} and weekly affect patterns \cite{stoneDayofweekMoodPatterns2012b}. We used the proportion of "good" or "somewhat good" responses from among all responses in a day as our independent variable, measuring an aggregate of mood per day. In total, we collected $268,128$ survey responses.

\subsection*{Sentiment analysis}

We applied the supervised GS classifier through the pytorch implementation distributed in the Hugging Face Hub (\url{https://huggingface.co/oliverguhr/german-sentiment-bert}). The underlying BERT model was trained on a diverse corpus to capture different types of expressions including social media text, reviews, Wikipedia, and newspaper articles in German. GS adds a sequence classifier head on top of the language model that is pretrained in a supervised fashion using three classes ("positive", "neutral", and "negative").

For the unsupervised approach, we use the German adaptation of the LIWC dictionaries \cite{wolf_computergestutzte_2008,pennebaker2015development}, in particular the word lists for positive emotions and for negative emotions. For efficiency reasons, we ran our own analyses \cite{pellertDashboardSentimentAustrian2020a} based on a version of the LIWC emotion dictionaries with small modifications to avoid systematic error on Twitter as in \cite{jaidkaEstimatingGeographicSubjective2020,metzler_collective_2021}. Specifically, we excluded COVID-19 related words (e.g. “treatment”, “heal”), and words that negatively correlate with happiness and well-being (love”, “good”, “LOL”, “better”, “well” and “like”) from the positive emotions dictionary. We tokenize the raw text of each post and compare the tokens to the LIWC dictionary categories, calculating a proportion of matching terms over all tokens in each post. 

To be able to combine different sentiment analysis methods, we rescaled measurements against baseline means calculated over a period preceding our analysis. We used the data corresponding to the first Austrian COVID-19 lockdown as a baseline (March 16th to April 20th 2020), since the period covered by the survey also corresponds to a lockdown in Austria. We rescaled daily sentiment aggregates by subtracting the baseline mean and dividing by it \cite{pellertDashboardSentimentAustrian2020a}.
To construct an aggregate of emotions comparable to the survey, we calculated the aggregate as the rescaled measure of positive emotions minus the rescaled measure of negative emotions. Our aggregate sentiment measure is the arithmetic mean of both methods GS and LIWC. As the survey question is targeted at feelings experienced on the previous day, responses can be influenced by feelings experienced on the day of the question and also two days before when a user responds just after midnight. To take this into account, we calculated all our sentiment aggregates on social media text over rolling windows of three days. We chose on purpose only one well-established dictionary-based method (LIWC) and one innovative, out-of-the-box method based on deep learning (GS). The pre-registered part of our analysis (\url{https://aspredicted.org/blind.php?x=vb3gp2}) demonstrates rigidly that we did not employ post-hoc additional methods out of the large pool available to influence our results \cite{chanFourBestPractices2021}. 

\subsection*{Statistical analysis}

We use the cocor R package \cite{diedenhofenCocorComprehensiveSolution2015,hittnerMonteCarloEvaluation2003} to statistically assess if two correlations (between sentiment aggregates or components and the survey) are statistically significantly different. Additionally, we perform bootstrap sampling of the differences between correlation coefficients and report the results in Figure S4. We fit models of our sentiment measures and the survey for both platforms as well as models of the changes of both sets of variables with "lm" in R \cite{rcite}. We use HAC correction in the R package "sandwich" \cite{sandwich1,sandwich2} to provide a robust assessment of the informativeness of Twitter and Der Standard signals when autocorrelation and heteroscedasticity might be present. $\beta$ coefficients that we report of our models and their confidence intervals are HAC corrected.

\subsection*{Text data}

Between March 6th and December 30th, 2020, our Austrian social media emotions monitor retrieved $4,161,820$ posts in German from the forum on derstandard.at \cite{pellertDashboardSentimentAustrian2020a}. All retrieved posts and all survey responses were considered in our analysis, i.e. there was no exclusion criterion. Our dataset includes $11,237$ unique users that published $635,185$ posts on derstandard in the subperiod that overlaps with the survey. For calculating baseline values, we used $1,021,978$ posts by $15,871$ users in the period between March 16th and April 20th, 2020. In the same period of March 6th and December 31st, 2020 we collected $5,886,805$ tweets from Twitter. We used information available about the number of followers to exclude users with more than 100,000 followers. This criterion and the shift of one day for the Twitter data are the only additional analyses with changes not included in the pre-registration \url{https://aspredicted.org/blind.php?x=vb3gp2} that follows the same methodology as for text data from derstandard.at. We used the archive of a data aggregation service (Brandwatch, formerly known as Crimson Hexagon) to retrieve all tweet IDs that were posted in Austria during the full period. The rehydration procedure that retrieves the full tweet object from the IDs makes sure that user's decision to remove their tweets from public display are respected. For the subperiod of the survey, our Twitter data set includes $11,237$ unique users with $635,185$ tweets on Twitter (same baseline period as for Der Standard using $743,003$ tweets by $11,082$ users).
To investigate possible differences between user postings on the two platforms, we compared them using word clouds on the days of two important events in Figure S12 \& S13. We find the focus of topics of discussions to be comparable. Similarly, comparing the two full text corpora for the entire survey period reveals no surprising differences in word frequencies (Figure S14).

\subsection*{Data Statement}

All data used in this research was either publicly available archival data (Twitter and Der Standard postings) or available to us as anonymized aggregated counts (survey data). The survey did not collect any individual or personally identifiable data. All data retrieval and analysis protocols comply with the regulations for ethical scientific practice at Graz University of Technology. We publish R scripts in a GitHub repository at \url{https://github.com/maxpel/SocialMediaMacroscopes} to replicate all figures and tables from this manuscript. We include daily aggregates of our measures for both platforms but do not redistribute text, as it could include personally identifying information and could be de-anonymized through search methods. For Der Standard forums, posts are publicly accessible on the newspaper site and can be retrieved respecting the rules of the platform. All posts that remain publicly available on Twitter can be retrieved for academic research by their IDs which are available in the repository.

\section*{Acknowledgements}

This work has been supported by the Vienna Science and Technology Fund (WWTF) through the project “Emotional Well-Being in the Digital Society” (Grant No. VRG16-005) and through project COV20-027.
We thank Thomas Niederkrotenthaler, Vibrant Emotional Health and the Vienna Science and Technology Fund for providing the funding to access Brandwatch. We also thank Thomas Niederkrotenthaler for providing demographic data from the representative Austrian survey, which we used to describe Austrian users of derstandard.at and Twitter. The authors declare that they have no competing interests. 

\section*{Author contributions statement}

D.G., M.P. conceived the research, M.M collected survey data, D.G., M.P. collected social media data, M.P. developed text analysis methods and performed data analysis, D.G., M.P., H.M. created preregistration, M.P., D.G. performed the statistical analysis. M.P., D.G. and H.M. wrote the manuscript. All authors reviewed and approved the manuscript.

\renewcommand{\thefigure}{S\arabic{figure}}
\renewcommand{\thetable}{S\arabic{table}}

\setcounter{figure}{0}
\setcounter{table}{0}

\clearpage

\section*{Supplementary Information}

\begin{figure*}[!htbp]
\centering
\includegraphics[width=0.8\linewidth]{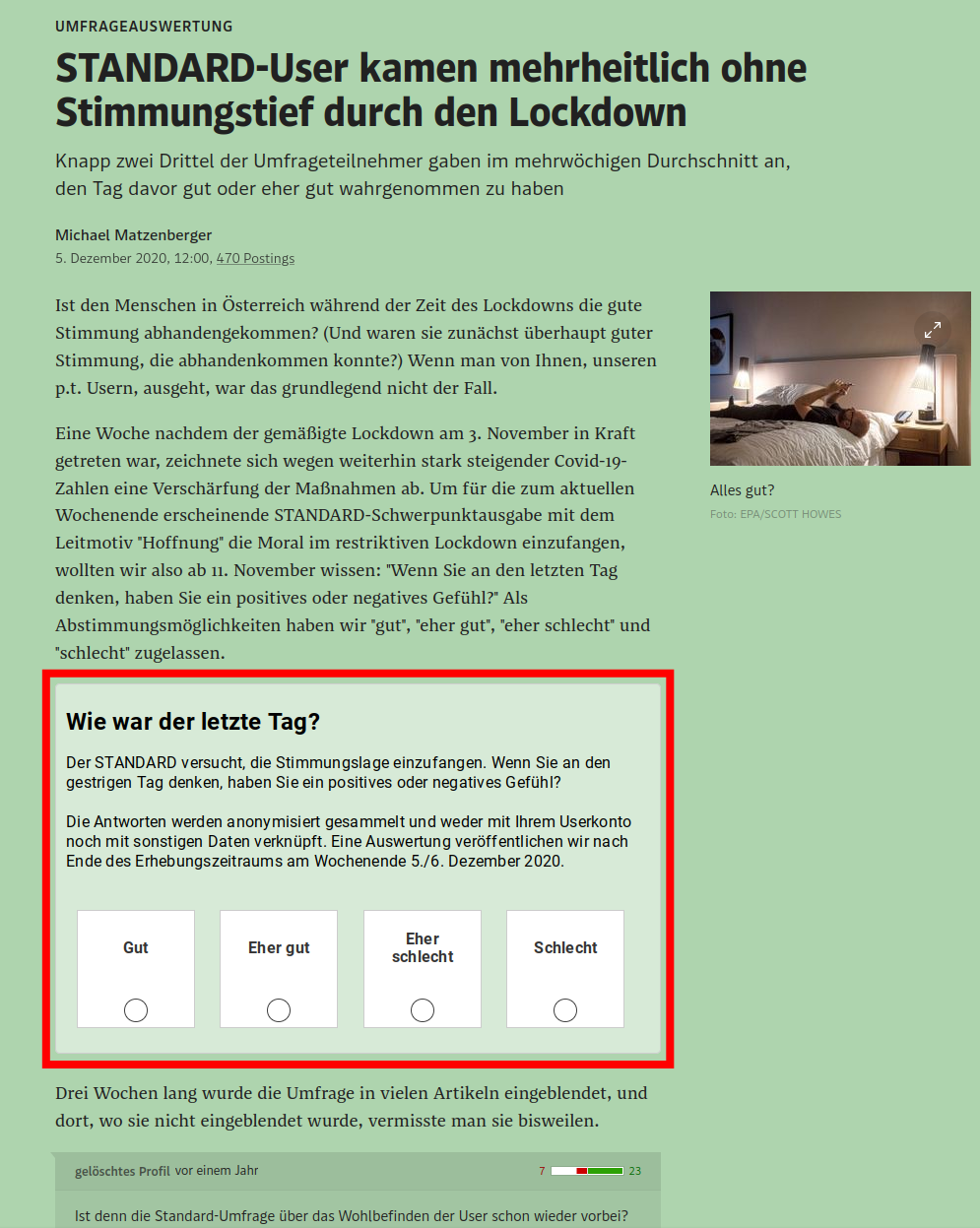}
\caption{Screenshot of derstandard.at showing a news article where the survey (red rectangle added by the authors) is displayed in between the article text. The full text of the survey contains the following information (translated to English by the authors): "How was the last day? Der Standard tries to capture general mood. If you think about yesterday, do you have a positive or negative feeling?
Answers are collected anonymously and not linked to your user account, nor any other data. The results will be published at the end of the survey period on the weekend of December 5/6 2020."}\label{fig:survey}
\end{figure*}

\begin{figure*}[!htbp]
\centering
\includegraphics[width=0.9\linewidth]{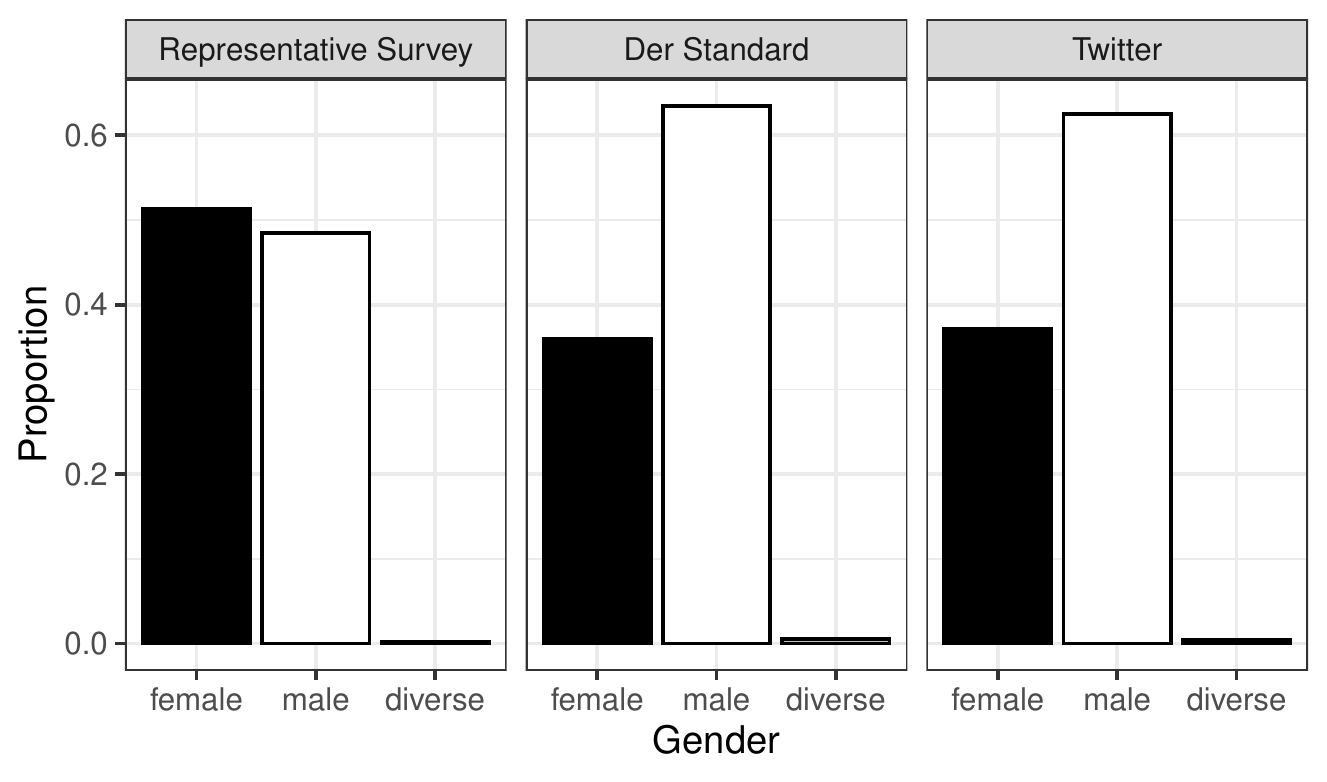}
\caption{Males are over-represented both on Der Standard and on Twitter, and to a similar extent.}\label{fig:demo_gender}
\end{figure*}

\begin{figure*}[!htbp]
\centering
\includegraphics[width=0.9\linewidth]{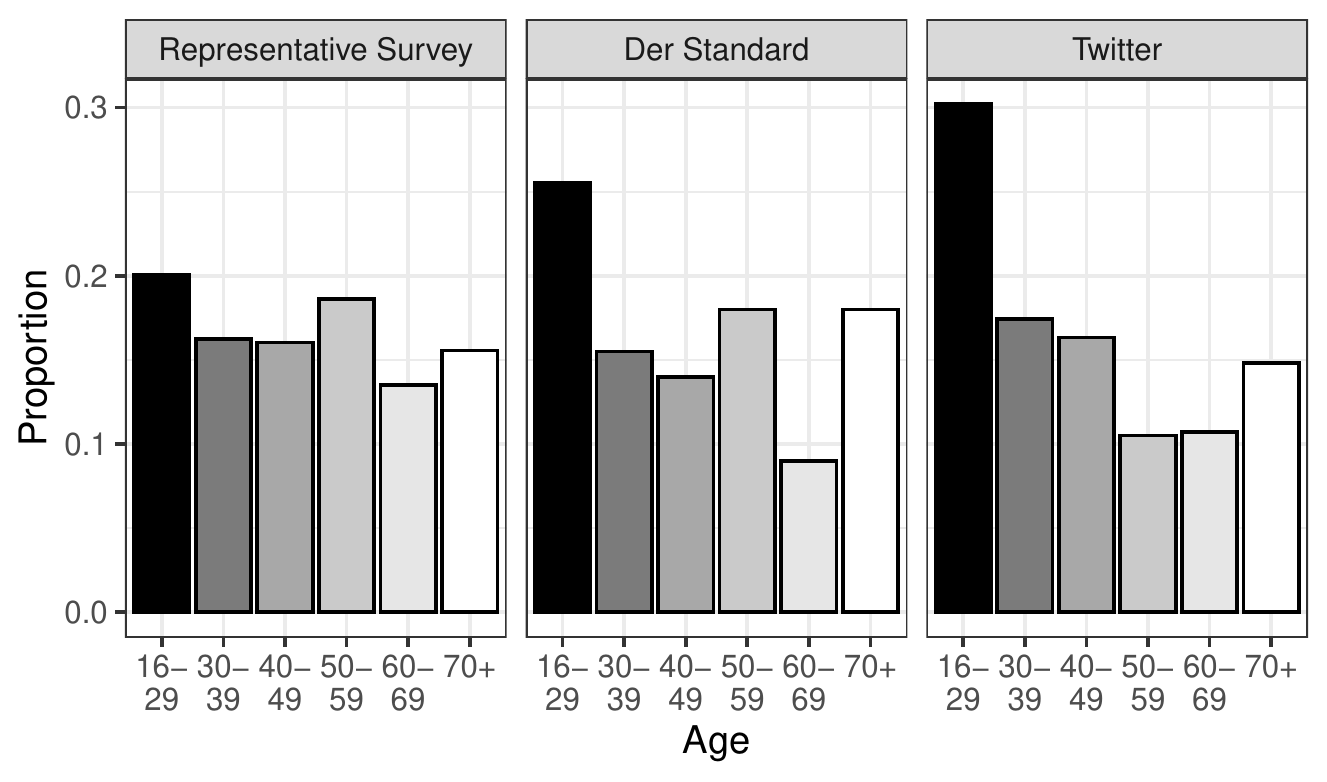}
\caption{The youngest cohort (16-29) is over-represented on Der Standard and even more so on Twitter. 50-59 year-olds are under-represented on Twitter only, whereas 60-69 year-olds are under-represented on both platforms.}\label{fig:demo_age}
\end{figure*}

\begin{figure*}[!htbp]
\centering
\includegraphics[width=0.9\linewidth]{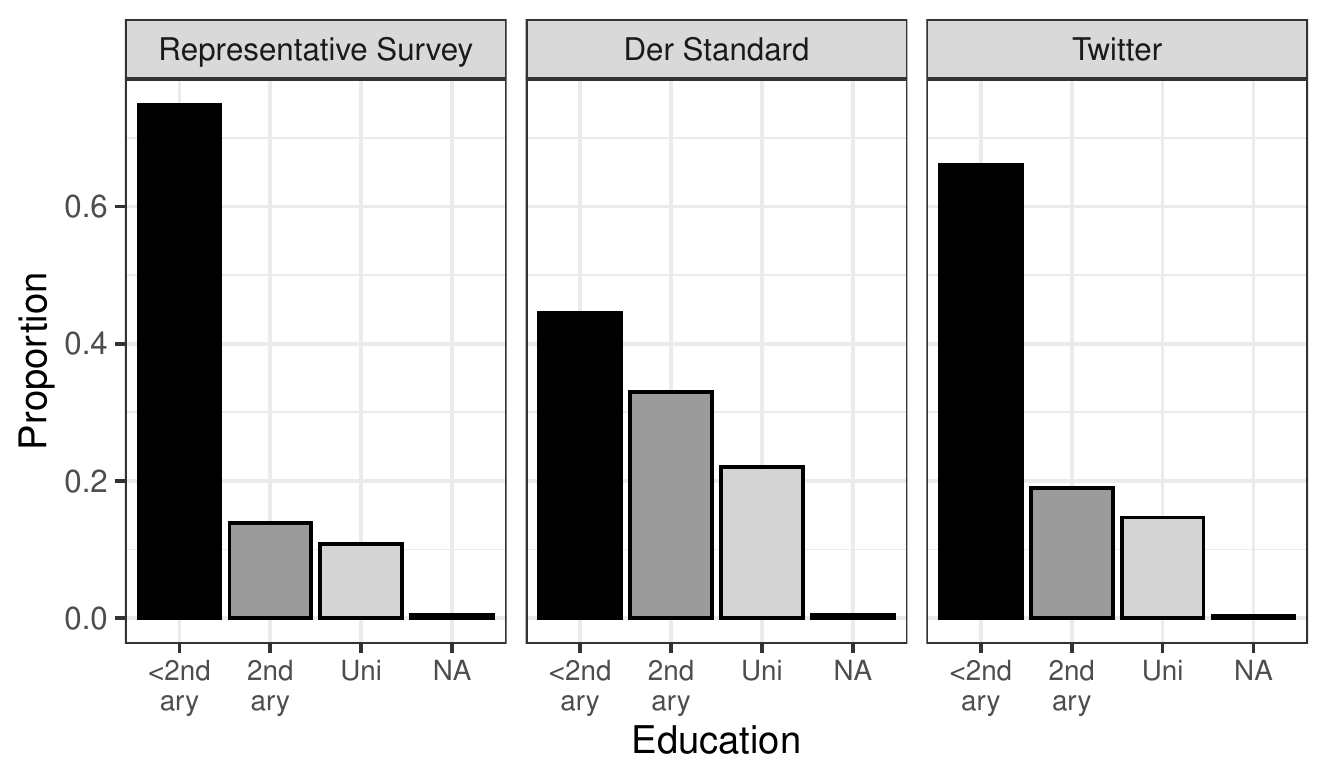}
\caption{Users with education below secondary school ("Matura" in Austria) are under-represented, especially on Der Standard, whereas higher education is over-represented.}\label{fig:demo_education}
\end{figure*}

\begin{figure*}[!htbp]
\centering
\includegraphics[width=0.9\linewidth]{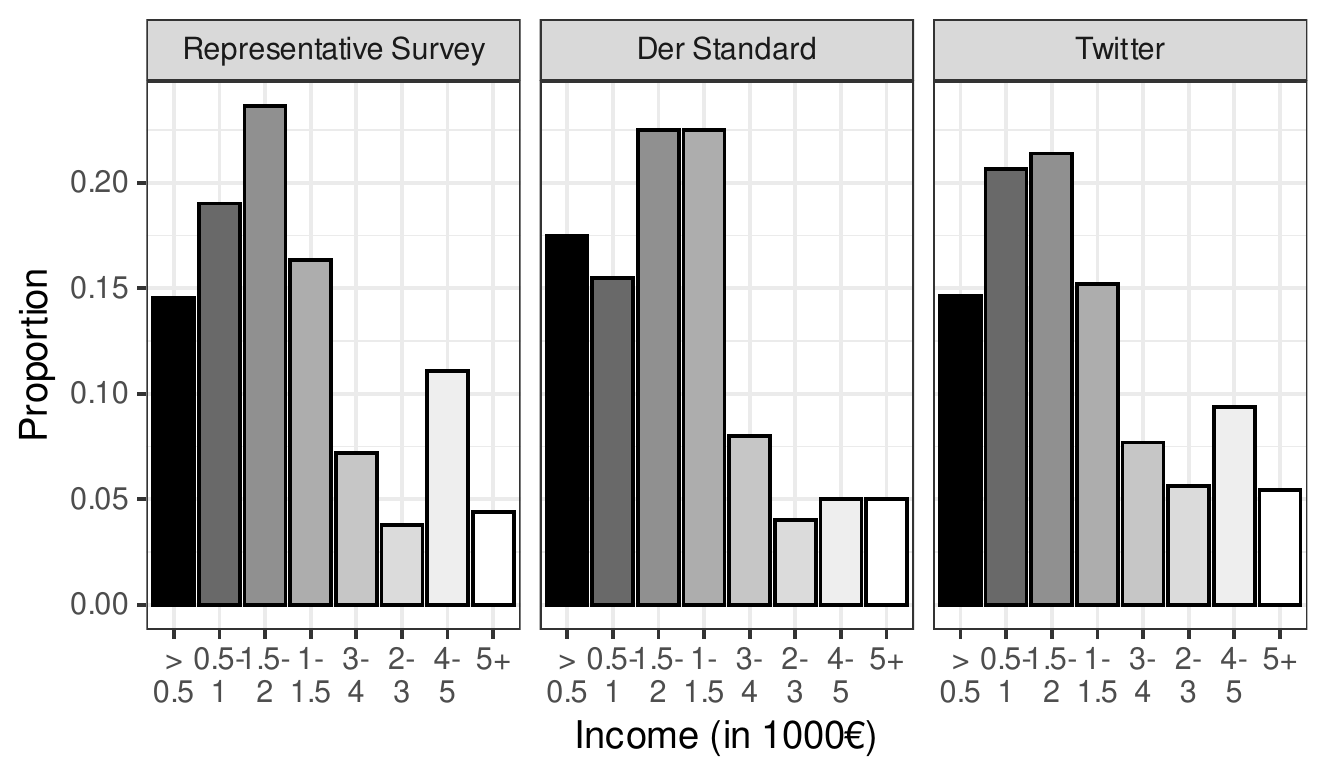}
\caption{Self-reported income does not differ strongly, although some categories might be slightly over- or underrepresented.}\label{fig:demo_income}
\end{figure*}

\begin{figure*}[!htbp]
\centering
\includegraphics[width=0.9\linewidth]{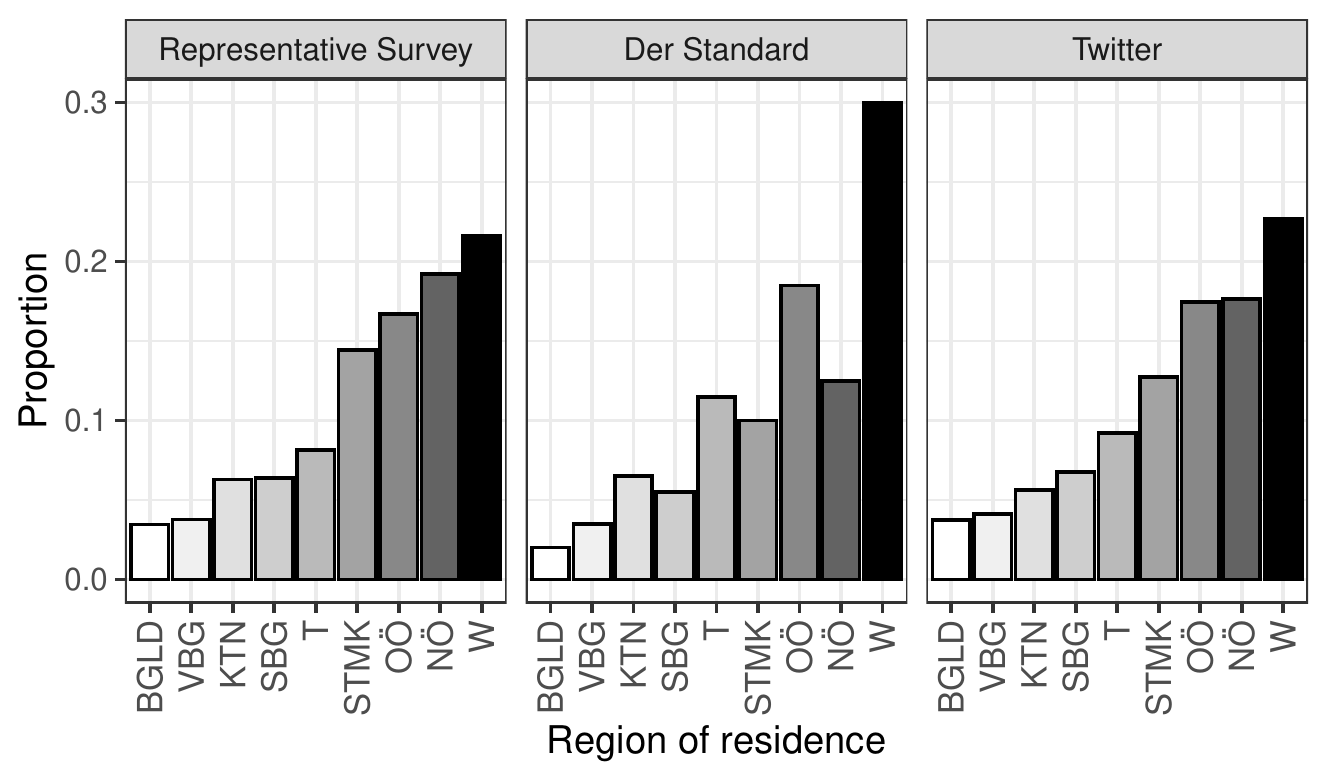}
\caption{Region of residence is close to representative on Twitter. On Der Standard, Vienna and Upper Austria ("OÖ") are over-represented, while Styria ("STMK") and Lower Austria ("NÖ") are under-represented.)}\label{fig:demo_region}
\end{figure*}

Table \ref{tab:diffcasecor} shows that for Der Standard, correlations of LIWC and aggregate sentiment with COVID-19 cases are significantly lower compared to the survey. In contrast, the correlation of GS sentiment alone is not significantly different compared to the survey. For Twitter, none of the correlations with COVID-19 cases differs between the sentiment measures and the survey. Overall, the signal from the survey is thus not more correlated with new cases than Twitter sentiment. Yet, when using data from Der Standard, only one of the sentiment methods (GS) correlates as strongly with cases as the survey, suggesting that the LIWC dictionary could be improved for this particular data source.

\begin{table}[!htbp]
\centering
\caption{Significance tests comparing correlations with the number of new COVID-19 cases between aggregate sentiment on both platforms and the survey. We test the difference in correlations of sentiment measures with new COVID-19 cases compared to the survey with new COVID-19 cases.}
\bigskip
\label{tab:diffcasecor}
\begin{tabular}{rrl}
  \toprule
 & p value & Difference (higher) \\ 
  \midrule
Twitter (Aggregate Sentiment) vs. Survey & 0.501 & 0.06 (Twitter) \\ 
  Twitter (LIWC) vs. Survey & 0.617 & 0.05 (Twitter LIWC) \\ 
  Twitter (GS) vs. Survey & 0.644 & 0.04 (Twitter GS) \\ 
  Der Standard (Aggregate Sentiment) vs. Survey & 0.015 & -0.2 (Survey) \\ 
  Der Standard (LIWC) vs. Survey & 0.023 & -0.36 (Survey) \\ 
  Der Standard (GS) vs. Survey & 0.087 & -0.16 (Survey) \\ 
   \bottomrule
\end{tabular}
\end{table}

\begin{figure*}[!htbp]
\centering
\includegraphics[width=0.98\linewidth]{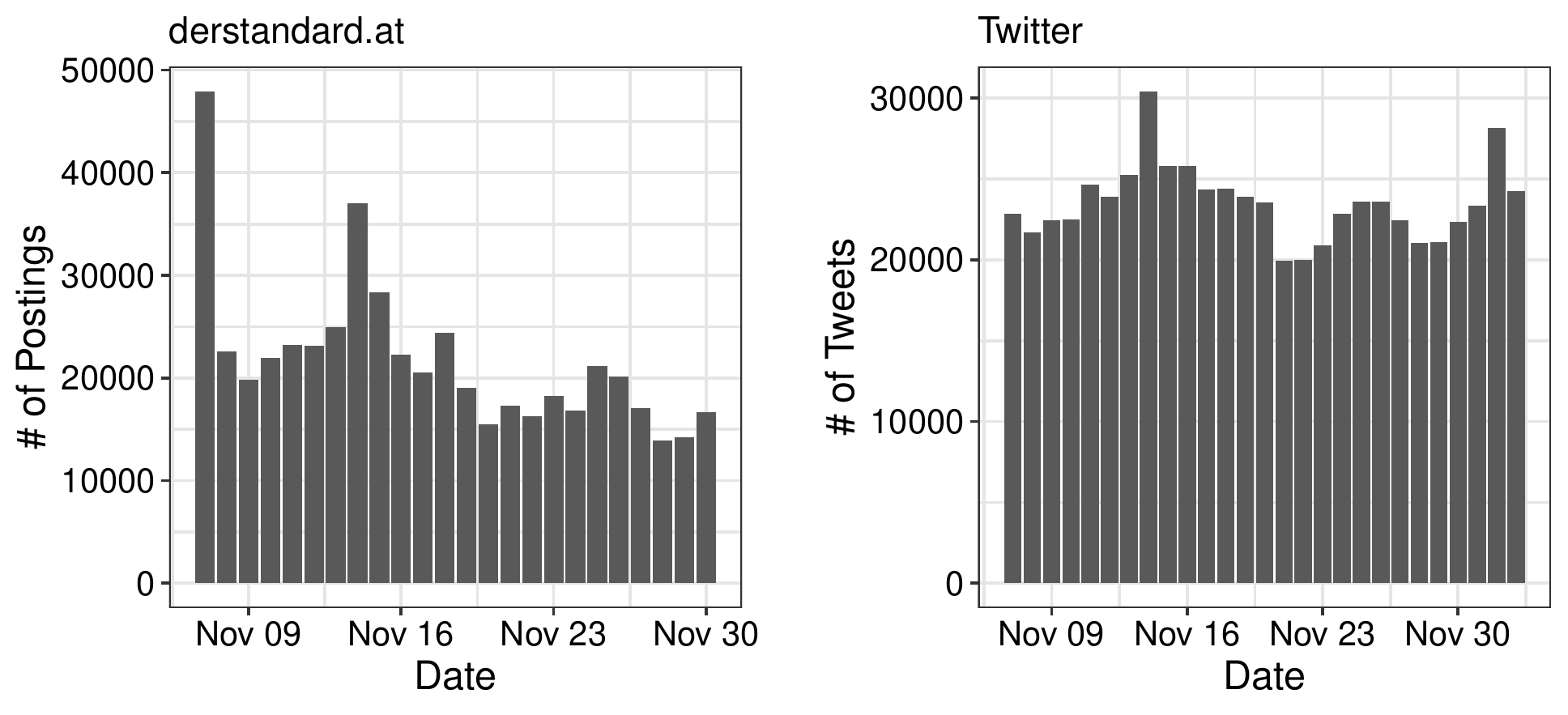}
\caption{Number of posts per day for both platforms.}\label{fig:nobs}
\end{figure*}

\begin{figure*}[!htbp]
\centering
\includegraphics[width=0.98\linewidth]{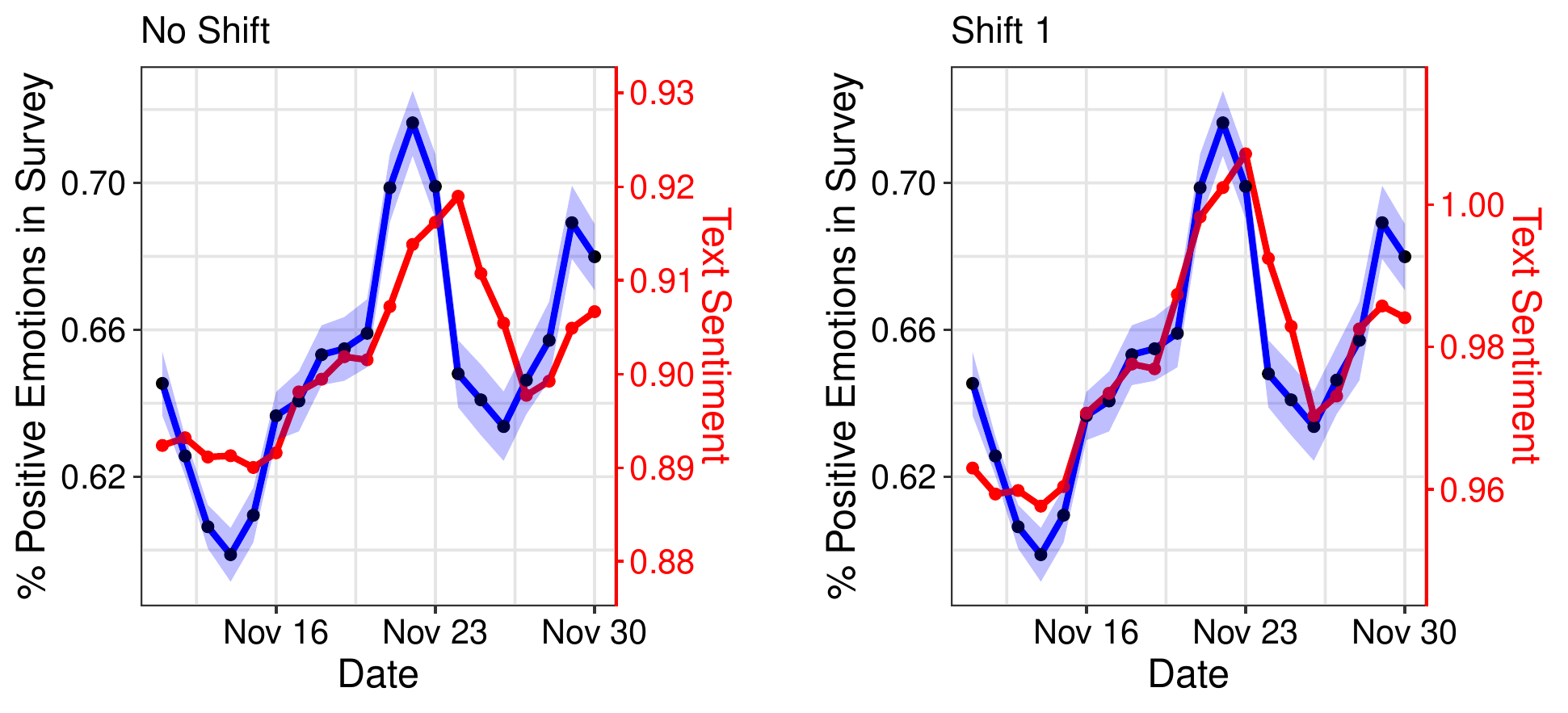}
\caption{Twitter is slightly delayed compared to postings on derstandard.at postings. A shift of one day corrects the slower response on Twitter.}\label{fig:shiftcompare}
\end{figure*}

\begin{figure*}[!htbp]
\centering
\includegraphics[width=0.98\linewidth]{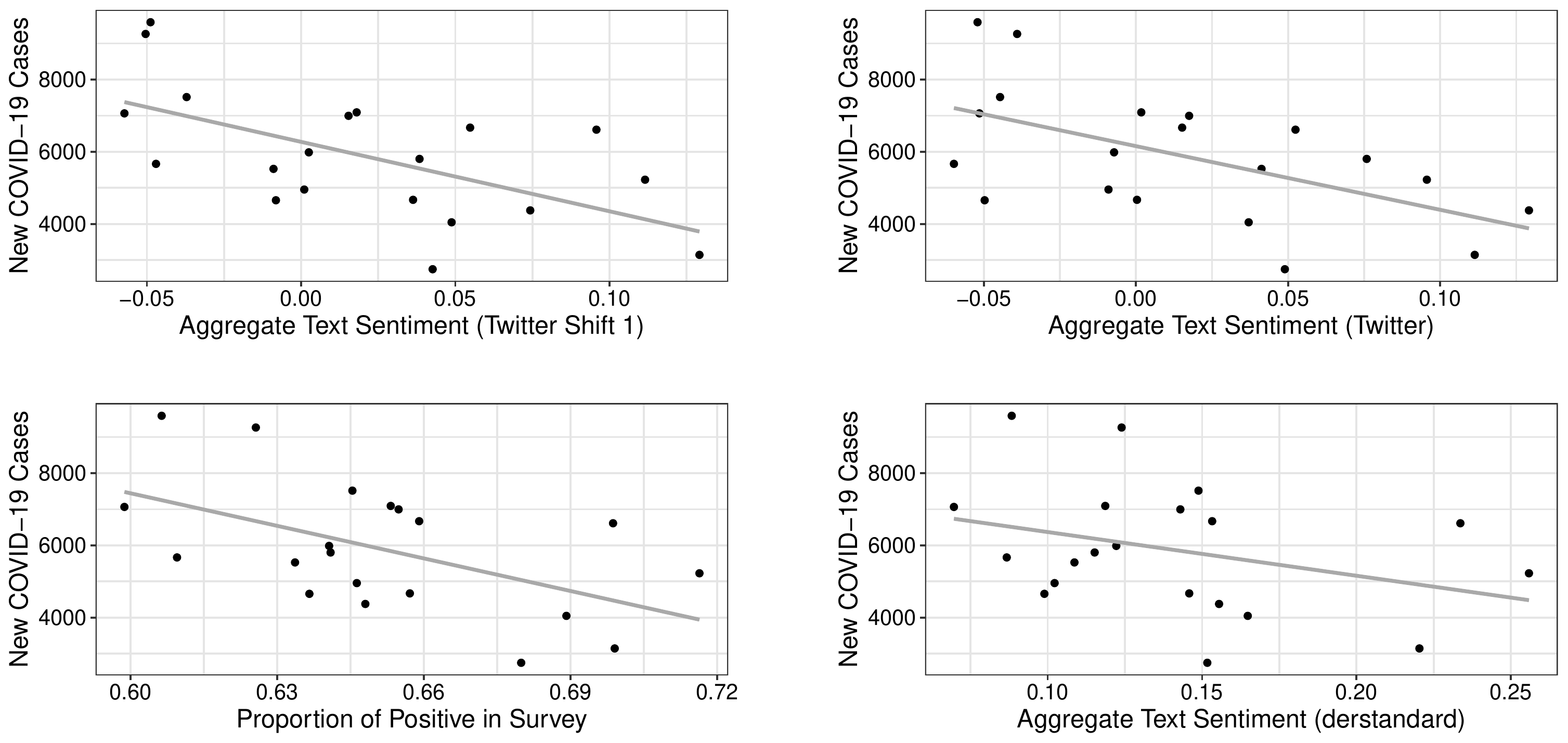}
\caption{Scatterplots for aggregate text sentiment and positive survey responses (see Table 3 in the main document) vs. the number of new COVID-19 cases in Austria.}\label{fig:scatter}
\end{figure*}

\begin{figure*}[!htbp]
\centering
\includegraphics[width=0.75\linewidth]{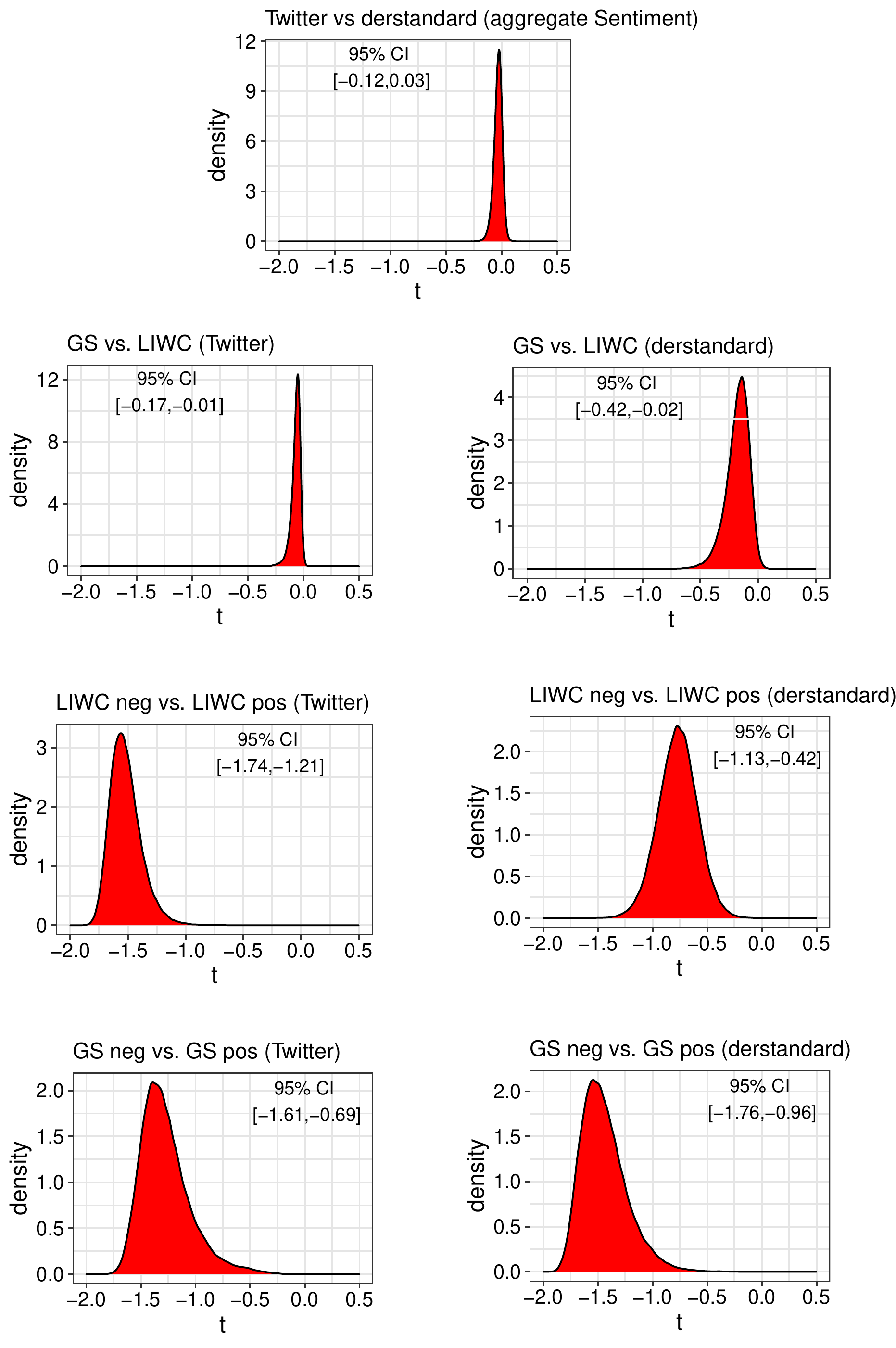}
\caption{Results of bootstrapping the differences between two correlation coefficients (Table 1 in the main document) for 100 000 times.}\label{fig:bootcompare}
\end{figure*}

\begin{figure*}[!htbp]
\centering
\includegraphics[width=0.95\linewidth]{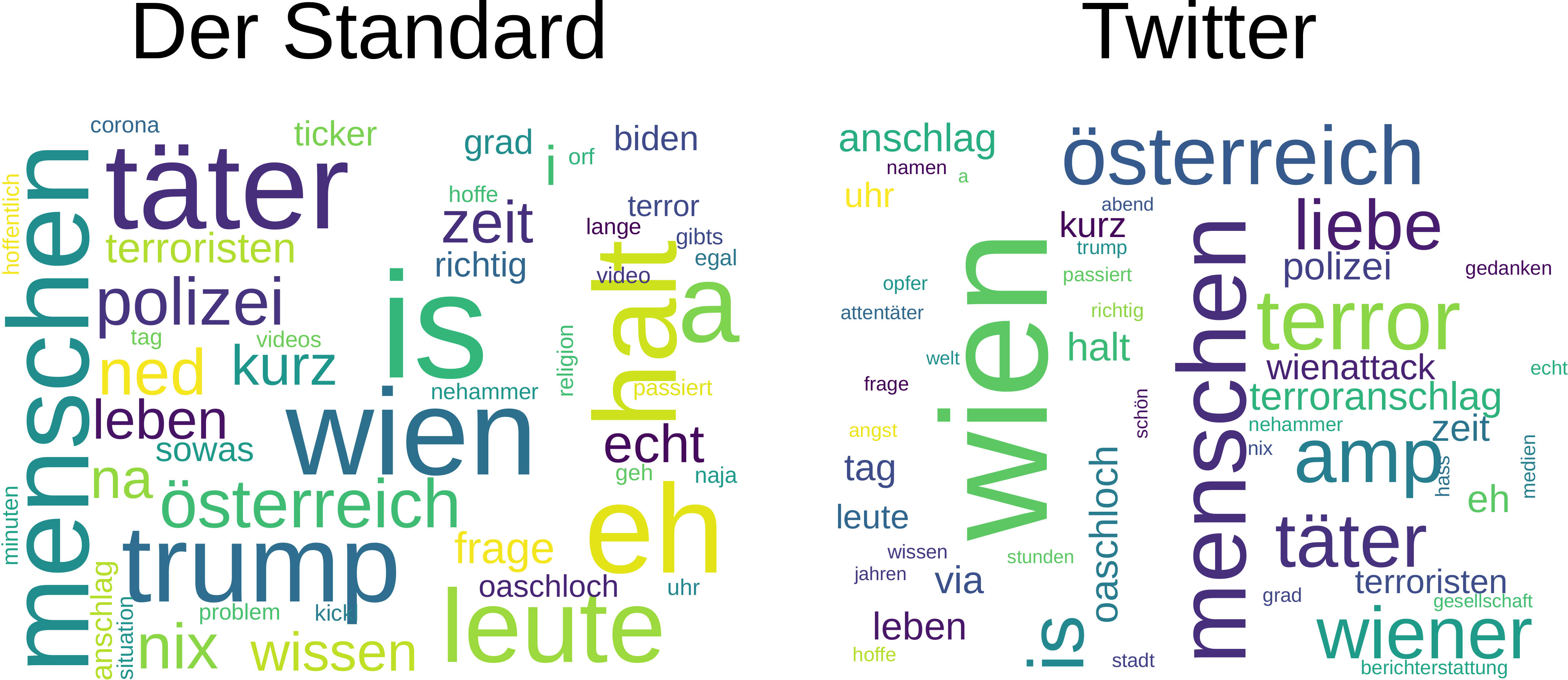}
\caption{Word clouds of the text on both platforms on 2020-11-03, the day after a terrorist attack in Vienna. We see references to "Täter"/perpetrator and location ("Wien"/Vienna), "Menschen"/"people" on both platforms. We display the 40 most common words and filter stop words using the full list of German stop words from \url{https://github.com/solariz/german_stopwords}.}\label{fig:wcterror}
\end{figure*}

\begin{figure*}[!htbp]
\centering
\includegraphics[width=0.95\linewidth]{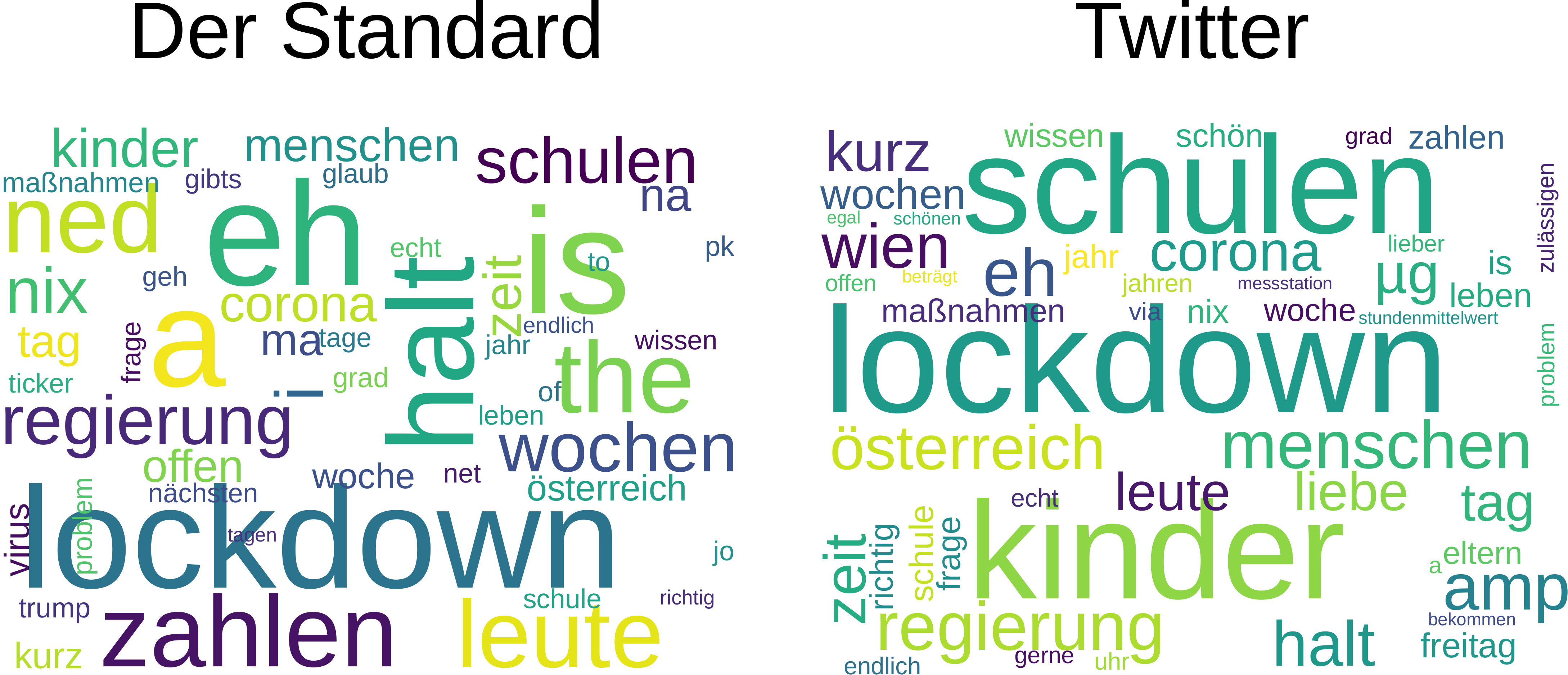}
\caption{Word clouds of the text on both platforms on 2020-11-13, the day of the highest increase of cases in the second COVID-19 wave in Austria. References to "lockdown", "kinder"/"children" and "schulen"/"schools" reflect similar discussions on both platforms. We display the 40 most common words and filter stop words using the full list of German stop words from \url{https://github.com/solariz/german_stopwords}.}\label{fig:wclarge}
\end{figure*}

\begin{figure*}[!htbp]
\centering
\includegraphics[width=0.5\linewidth]{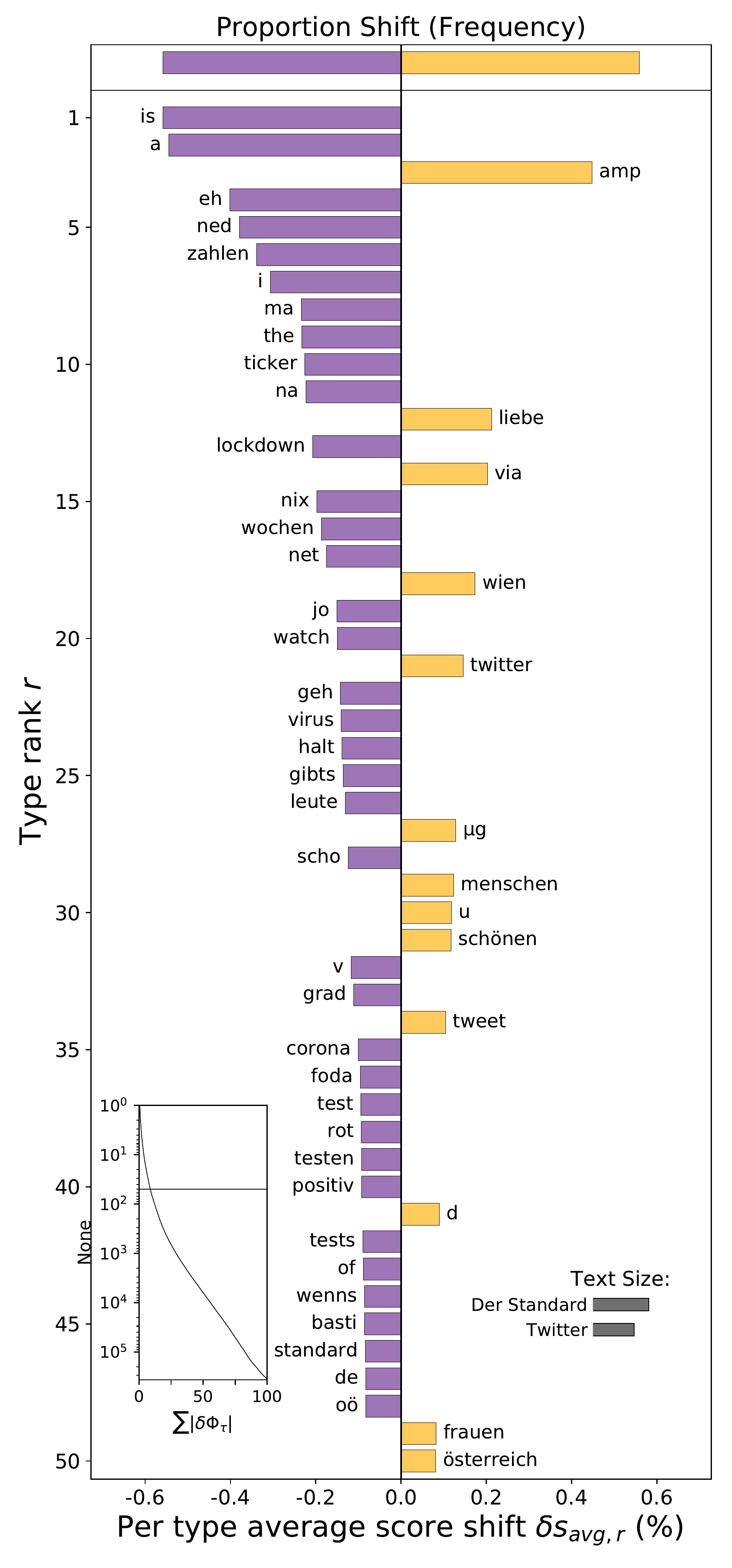}
\caption{Wordshift graph comparing the text of both platforms during the survey period from 2020-11-11 - 2020-11-30. The shift shows Twitter minus Der Standard. The purple (yellow) bars show words that are more frequent on Der Standard (Twitter). The biggest differences in the corpora are trivial dialect words of Austrian German on Der Standard. Apart from that, there seem to be more direct mentions of COVID-19 related concepts (virus, test, lockdown, test) on Der Standard. We filter stop words using the full list of German stop words from \url{https://github.com/solariz/german_stopwords}.}\label{fig:wsgraph}
\end{figure*}

\clearpage

\bibliography{scibib}

\bibliographystyle{Science}

\end{document}